# MgPd$_2$Sb – the first Mg-based Heusler-type superconductor

M.J. Winiarski[1,2] †, G. Kuderowicz[3], K. Górnicka[1,2], L.S. Litzbarski[1,2],
K. Stolecka[1], B. Wiendlocha[3], R.J. Cava[4], T. Klimczuk[1,2] $

[1]*Faculty of Applied Physics and Mathematics,
Gdansk University of Technology, Narutowicza 11/12, 80-233 Gdansk, Poland*

[2]*Advanced Materials Centre,
Gdansk University of Technology, ul. Narutowicza 11/12, 80-233 Gdańsk, Poland,*

[3]*Faculty of Physics and Applied Computer Science,
AGH University of Science and Technology, Aleja Mickiewicza 30, 30-059 Kraków, Poland*

[4]*Department of Chemistry, Princeton University, Princeton, NJ 08544, USA*

† michal.winiarski@pg.edu.pl
$ tomasz.klimczuk@pg.edu.pl

**Abstract**

We report the synthesis and physical properties of a new full Heusler compound, MgPd$_2$Sb, which we found to show superconductivity below $T_c$ = 2.2 K. MgPd$_2$Sb was obtained by a two-step solid state reaction method and its purity and cubic crystal structure (F m-3m, a = 6.4523(1) Å) were confirmed by powder x-ray diffraction. Normal and superconducting states were studied by electrical resistivity, magnetic susceptibility and heat capacity measurements. The results show that MgPd$_2$Sb is a type-II, weak coupling superconductor ($\lambda_{ep}$ = 0.56). The observed pressure dependence of $T_c$ ($\Delta T_c/p \approx$ -0.23 K/GPa) is one of the strongest reported for a superconducting Heusler compound. The electronic structure, phonons and electron-phonon coupling in MgPd$_2$Sb were theoretically investigated. The obtained results are in agreement with the experiment, confirming the electron-phonon coupling mechanism of superconductivity. We compare the superconducting parameters to those of all reported Heusler-type superconductors.

## Introduction

Heusler phases constitute a large family of intermetallic compounds with hundreds of members [1,2] The prototype member, MnCu$_2$Al was discovered by Fritz Heusler in 1903 and was the first known ferromagnetic material that was not composed of elements showing ferromagnetism in their pristine form [3]. The cubic MnCu$_2$Al-type structure (space group no. 225, Fm-3m) can be considered as composed of a NaCl-type sublattice of *AB* (CuAl in the prototypic phase) with *C* (Mn) atoms filling all its tetrahedral voids. If only half of the voids are filled in an ordered manner, the so-called half-Heusler noncentrosymmetric (s.g.# 214, F-43m) structure results. In general, in most of the known Heusler compounds the *A* site is occupied by an electropositive element (rare-earth metals, early transition elements), *B* by an electronegative *p*-block element (groups 13-15 of the periodic table) and the *C* site by a late transition metal (groups 9-11).

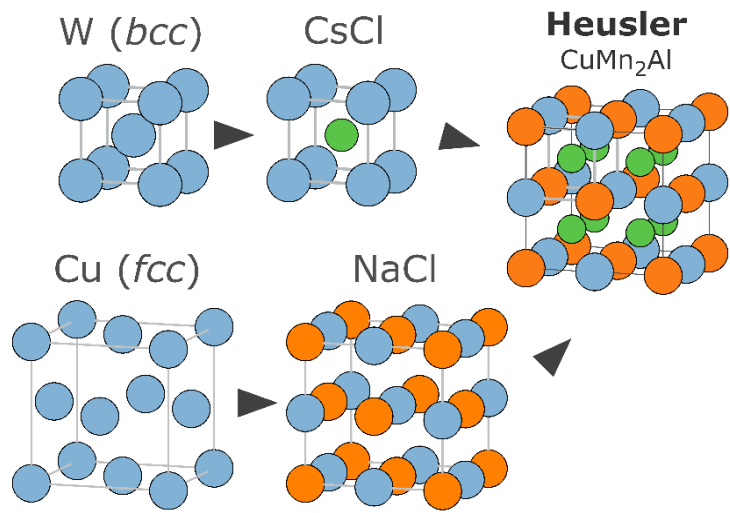

**Fig. 1 Relationship between the (full) Heusler and simple close-packed structures. The Heusler structure (e.g. of CuMn$_2$Al) can be viewed as an ordered superstructure of CsCl-type structure, which itself is an ordered variant of a simple *bcc* lattice as found in elemental W. Alternatively the Heusler structure can be obtained by filling tetrahedral voids of the NaCl-type lattice, which is an ordered variant of the *fcc* cell, represented eg. by metallic Cu. In half Heusler phases (MgAgAs type) only every other tetrahedral void of the NaCl-substructure is filled, resulting in a noncentrosymmetric structure.**

Both Heusler and half-Heusler phases have attracted significant research interest, due to their chemical versatility leading to a plethora of ground states and properties, including superconductivity [4–8], topologically non-trivial states [9–13], and half-metallic ferromagnetism [14–18]. Many of the properties can be tuned by chemical substitution [16,19–21] or mechanical strain [22,23], and the relative chemical and structural simplicity allows the growth of Heusler phase-based superstructures and nanosized (thin films, nanowires) forms [22–25].

Only 34 full Heusler type superconductors are reported, with critical temperatures reaching up to $T_c \approx$ 5 K (YPd$_2$Sn) [4,5], all of them featuring rare-earth elements or early transition metals (Zr, Hf, Nb) at the *A* site and most of them having Pd at the *C* site [5,8]. A list of selected superconducting parameters for for all members of this family of superconductors can be found in Table 1.

Based on a previously observed relation between the superconducting $T_c$ and the valence electron count (VEC), showing a maximum around VEC = 27 [5], we have selected MgPd$_2$Sb as a candidate superconductor. Although not previously reported experimentally, the compound was predicted to be stable based on high-throughput electronic structure calculations in the Materials Project database [26,27]. Thanks to their simple crystal structure, total energies of numerous hypothetical Heusler-type compounds were calculated, which allows prediction of their thermodynamic stability. It is worth noting that in the Open Quantum Materials Database, OQMD, the Heusler phase of MgPd$_2$Sb (id:468110) is predicted to be slightly less stable than tetragonal (id:1014836) and orthorhombic (id:1015910) variants of the same composition, with a minute energy difference of ca. 2 meV/atom [28,29]). In this paper we report superconductivity in MgPd$_2$Sb and discuss its properties in the context of the Pd-based Heusler superconductor family. Our finding supports the applicability of the VEC-T$_c$ relationship of Heusler superconductors.

**Materials and methods**

Polycrystalline samples of MgPd$_2$Sb were synthesized via a two-step solid state reaction. First a precursor compound, PdSb, was synthesized by reacting Pd and Sb powders pressed into pellet and annealed at 800°C for 5 h, sealed in an evacuated silica ampoule. Stoichiometric amounts of Pd and PdSb powders with a slight (2%) excess of metallic Mg flakes were mixed, pressed into pellets, wrapped with Ta foil and sealed in evacuated silica ampoules, backfilled with Ar to reduce the Mg evaporation during annealing. Ampoules were subsequently slowly heated (50°C/h) to 600°C, kept for 5 h, and furnace-cooled to room temperature. After the first annealing the resulting pellets were ground, re-pressed and annealed at 700°C for 5 h, resulting in relatively hard, dense samples. It is worth noting that similar to other Mg-based superconductors, i.e. MgCNi$_3$ [30] and Mg$_{10}$Ir$_{19}$B$_{16}$ [31], the Mg excess is important for observing superconductivity. However, we found that if more than 5-10% Mg surplus is taken, superconductivity is not observed and the full Heusler MgPd$_2$Sb decomposes to a half-Heusler MgPdSb structure.

Analysis of crystal structure of the synthesized samples was done by means of powder x-ray diffraction (XRD) using Bruker D2 Phaser diffractometer equipped with solid-state detector, using CuK$_\alpha$ radiation. XRD patterns were processed by means of the Rietveld refinement method using the FullProf software package [32].

Measurements of physical properties were performed using Quantum Design (QD) Physical Properties Measurement System (PPMS). For magnetization the vibrating sample magnetometer (VSM) option was used. Heat capacity was measured employing the standard semiadiabatic pulse technique in the He3 PPMS setup. The sample was attached to the measuring stage using Apiezon N grease to ensure good thermal contact. Resistivity measurements were performed using a four probe technique with thin Pt wires (d = 50 μm) mounted using silver epoxy to the surface of a rectangular bar cut out of the sample pellet. High-pressure resistivity was measured using a Quantum Design piston-cylinder-type pressure cell. Idemitsu Kosan Co. Daphne 7373 oil was applied as the pressure transmitting medium.

Following the experimental study, electronic structure, the phonons and electron-phonon coupling in MgPd$_2$Sb were theoretically investigated. Calculations were performed with the density functional theory implemented in Quantum Espresso [33,34] which uses the plane-wave pseudopotential method. Ultrasoft pseudopotentials were chosen [35] with the Perdew-Burke-Ernzerhof generalized gradient approximation exchange-correlation functional [36]. Computation details are: 60 Ry wavefunction energy cutoff, 600 Ry charge density cutoff, 24$^3$ k-point Monkhorst-Pack grid for the electronic structure and 8$^3$ q-point grid for the force constant matrix evaluation. The role of spin orbit coupling (SOC) was checked by including spin-orbit coupling and relativistic versions of pseudopotentials of Pd and Sb (replacing scalar Mg pseudopotential with the relativistic one had negligible effect) in the electronic structure and few selected q-vectors in phonon frequency calculations.

The unit cell was relaxed with the Broyden-Fletcher-Goldfarb-Shanno algorithm. Atomic positions are fixed by the symmetry, and hence only the lattice constant was varied. The calculated value for $a$, 6.548 Å, is close to the one calculated by the Materials Project [25] (6.553 Å), slightly larger than experimental one (see below). Then, the electronic bands, density of states (DOS) and Fermi surface were calculated. Phonons and their linewidths were calculated with the density functional perturbation theory [37]. Superconductivity with electron-phonon coupling was analyzed using Eliashberg functions, calculated from the phonon linewidths.

**Results and discussion**

Figure 1 shows the room temperature powder XRD pattern of MgPd$_2$Sb with a Rietveld fit. Resulting unit cell and structural parameters are gathered in Table 2. The refined unit cell parameter $a$ = 6.452 Å is in agreement with the above-mentioned computed values. No tetragonal or orthorhombic distortion, as predicted by OQMD calculations[27,28], was observed at room temperature within the experimental resolution of PXRD.

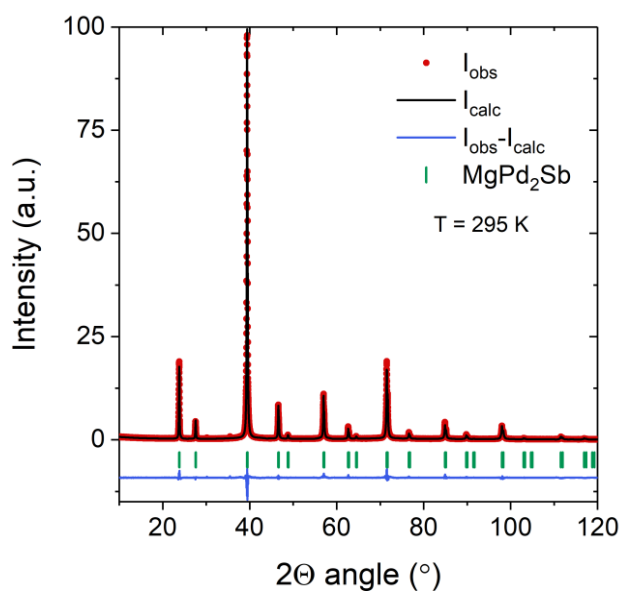

Figure 1 Rietveld fit (black line) to the room temperature powder XRD pattern of MgPd$_2$Sb (red points). Blue line shows the difference between observed and model intensities. Green ticks mark the expected positions of Bragg reflections.

Table 2 Unit cell parameters and isotropic thermal displacement factors derived from Rietveld fit to room temperature powder XRD data. Numbers in parentheses are statistical standard uncertainties of least significant digits.

| Formula unit: | MgPd$_2$Sb |
|---|---|
| $Z$ | 4 |
| Space group | F m -3 m (#225) |
| Unit cell parameter $a$ (Å) | 6.4523(1) |
| Cell volume $V$ (Å$^3$) | 268.62(1) |
| Calculated density $\rho$ (g/cm$^3$) | 8.87 |
| Reliability factors for Rietveld model: | $R_p$ = 8.73% |
|  | $R_{wp}$ = 10.3% |
|  | $R_{exp}$ = 4.92% |
|  | $\chi^2$ = 4.41 |
| Atomic positions | Isotropic thermal factors $B_{iso}$ (Å$^2$) |
| Mg (4$a$) (0 0 0) | 2.39(6) |
| Pd (8$c$) (¼ ¼ ¼) | 3.31(2) |
| Sb (4$b$) (0 0 ½) | 2.57(2) |

The temperature-dependent electrical resistivity of MgPd$_2$Sb (Fig. 1(a)) shows a typical metallic character with residual resistivity ratio RRR ≈ 1.9, a value similar to that observed in other Heusler superconductors [4]. A drop to zero is observed with an onset at ca. T = 2.2 K in zero magnetic field (Fig. 2(b)). The transition is slightly broad ($\Delta T_c$ ≈ 0.15 K, $\Delta T_c/T_c$ ≈ 7%), which may result from the presence of possible structural disorder. This suggests the presence of the antisite disorder commonly observed in Heusler compounds [4,30,31].

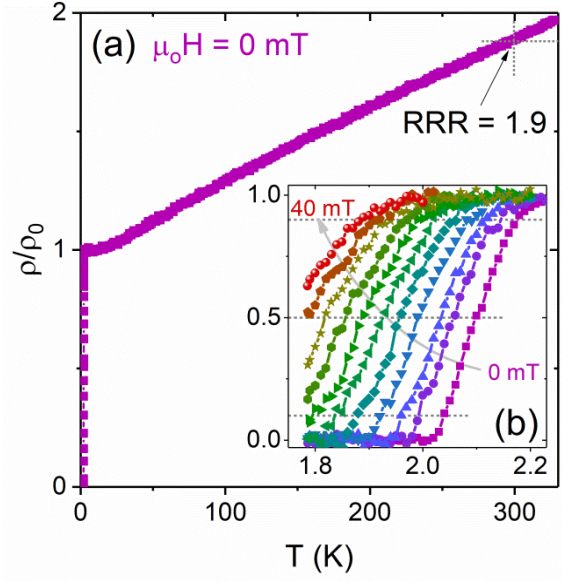

Figure 2 Normalized resistivity of MgPd$_2$Sb ($\rho(T)$ divided by residual resistivity $\rho_0$). Panel (a) shows $\rho(T)/\rho_0$ at zero field in the temperature range of 2-330 K. Inset (b): low-temperature resistivity in applied magnetic fields $\mu_0 H$ from 0 to 40 mT (5 mT spacing). Three horizontal dotted lines are drawn at 90%, 50% and 10% of the residual resistivity $\rho_0$.

Figure 3 shows the temperature dependence of the upper critical field as extracted from the resistivity measurement. A linear fit to the data yields the coefficient $d\mu_0 H_{c2}/dT$ = -144(2) mT/K and the zero-field $T_c$ = 2.1 K. Using the Werthamer-Helfand-Hohenberg relation [38,39]: the upper critical field $\mu_0 H_{c2}(0)$ can be estimated from:

$$\mu_0 H_{c2}(0) = -A T_c \left| \frac{d\mu_0 H_{c2}(T)}{dT} \right|_{T \to T_c}$$

Assuming the dirty limit value of the prefactor $A$ (0.69) and $T_c$ = 2.1 K the $\mu_0 H_{c2}(0)$ = 195 mT. The superconducting coherence length can be calculated using the extrapolated value of upper critical field via:

$$\xi_{GL} = \sqrt{\frac{\phi_0}{2\pi \mu_0 H_{c2}}}$$

Where $\phi_0$ is the quantum of magnetic flux. This gives $\xi_{GL}$ = 410 Å.

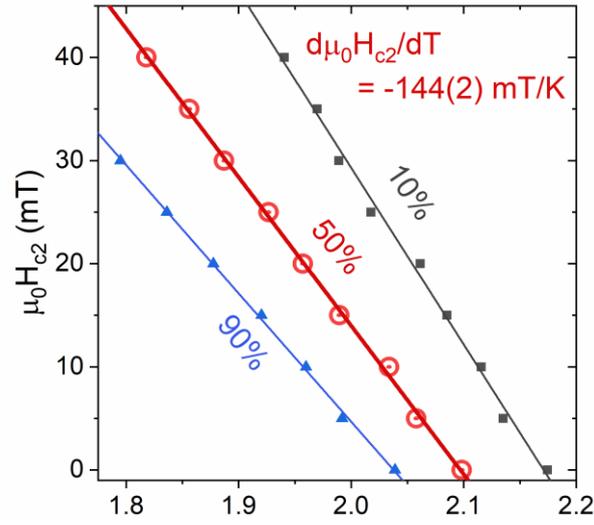

Figure 3 Temperature dependence of the upper critical field as extracted from temperature-dependent resistivity measurements at applied magnetic fields (see Fig. 2(b)) using 10%, 50%, and 90% resistivity drop criterion (gray, red, and blue points, respectively). Thick red line is a linear fit to the 50% data, yielding $d\mu_0H_{c2}/dT = -144(2)$ mT/K. The thin gray and blue lines are guides for eyes.

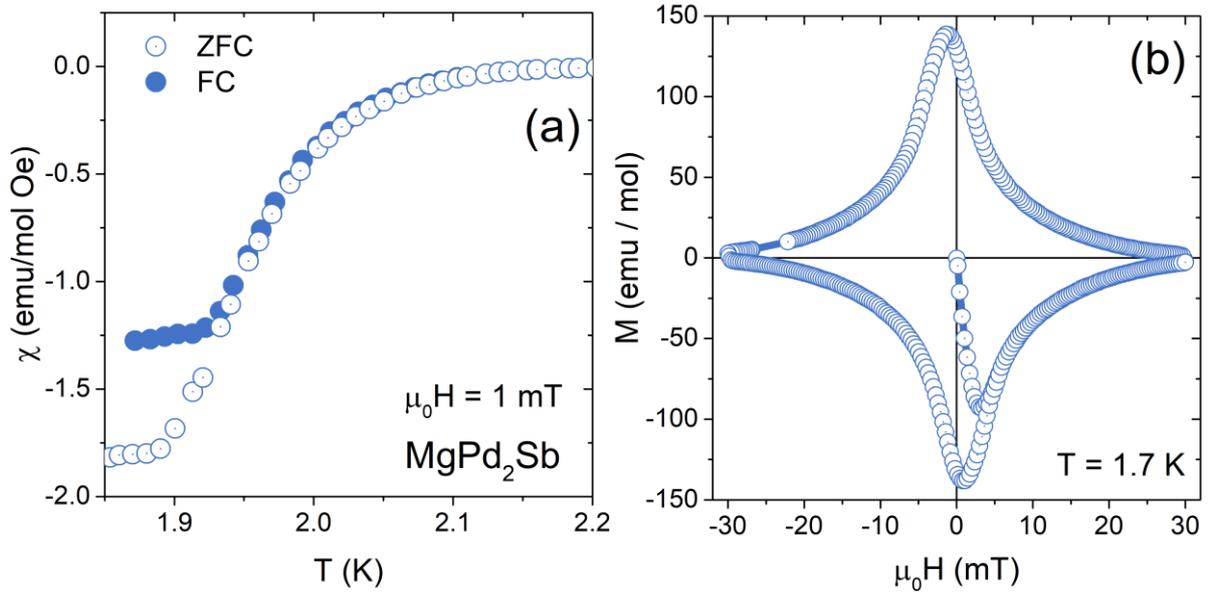

Figure 4 Panel (a): zero-field cooling (ZFC) and field-cooling (FC) *dc* molar magnetic susceptibility of MgPd$_2$Sb in an applied field of $\mu_0H = 1$ mT showing a transition to the Meissner state. Panel (b): hysteresis loop of the molar magnetization M(H) at 1.7 K.

To further characterize the superconducting state of MgPd$_2$Sb, we have performed measurements of the dc magnetization versus temperature, as well as versus magnetic field. Temperature-dependent *dc* magnetic susceptibility measured at applied magnetic field of H = 10 Oe is shown in Fig. 4(a). At T ≈ 2.15 K the onset of transition to the Meissner state is observed, with the critical temperature $T_c$ = 2.0 K. The difference between zero-field-cooled and field-cooled curves results from magnetic flux pinning between polycrystalline grains. Panel (b) presents the hysteretic M(H) loop measured at 1.7 K. The shape of the M(H) curves is suggestive of type-II superconductivity.

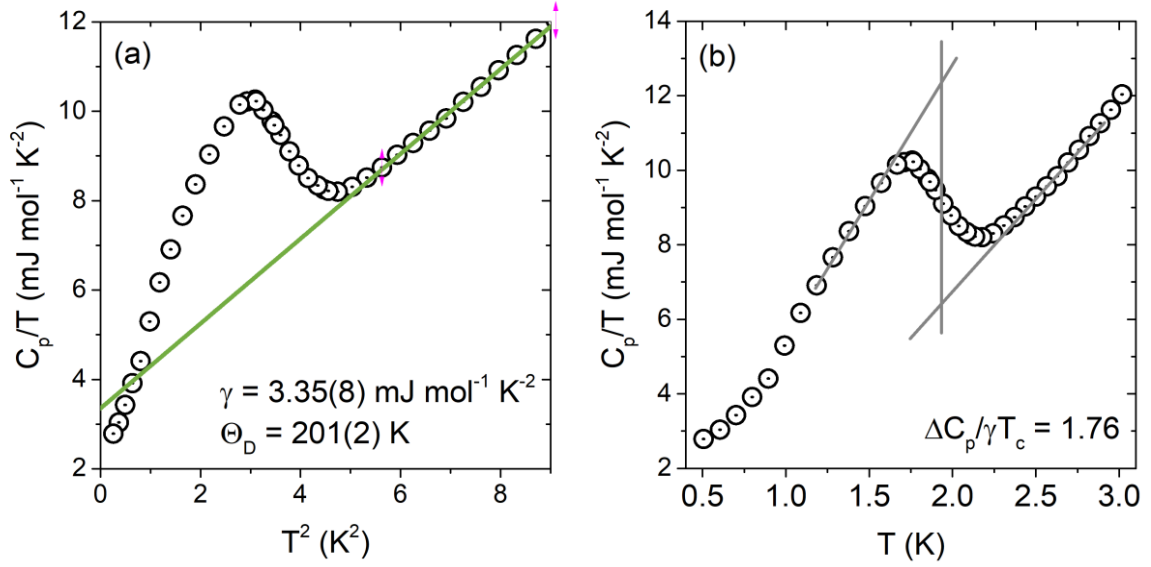

Fig. 5 Low-temperature heat capacity of MgPd$_2$Sb. Panel (a) shows a linear fit (green line) to the $C_p/T$ vs. $T^2$ data, yielding the γ and β electronic and lattice heat capacity coefficients. The Debye temperature calculated from β is $\Theta_D$ = 201 K. Panel (b) shows the equal entropy construction (gray lines) used to estimate the normalized heat capacity jump $\Delta C_p/\gamma T_c$ = 1.76.

The specific heat $C_p(T)$ between 0.5 K and 3 K measured in zero applied magnetic field is shown in Figure 5. A sharp lambda-shaped anomaly is seen with the transition temperature T = 1.95 K which was obtained by using an equal entropy construction (Fig. 5b). The $C_p(T)$ data above the transition temperature was fitted with the formula:

$$\frac{C_p}{T} = \gamma + \beta T^2$$

where γ is the Sommerfeld coefficient and β describes the phonon heat capacity contribution. The fit, shown by a solid line in Figure 5a, yields γ = 3.35(8) mJ mol$^{-1}$ K$^{-2}$ and β = 0.949(4) mJ mol$^{-1}$ K$^{-4}$. The β coefficient is related to the Debye temperature via:

$$\Theta_D = \sqrt[3]{\frac{12\pi^4 nR}{5\beta}}$$

The estimated $\Theta_D$ = 201 K falls within the range observed for other Heusler superconductors (180-240 K in $M$Pd$_2X$, $M$ = Sc, Y, Zr, Hf; $X$ = Al, In, Sn [4]).

Taking the $\Theta_D$ and $T_c$ values estimated from the heat capacity analysis the electron-phonon coupling parameter can be estimated using the inverted McMillan formula [40]:

$$\lambda_{ep} = \frac{1.04 + \mu^* \ln\left(\frac{\Theta_D}{1.45 T_c}\right)}{(1 - 0.62\mu^*) \ln\left(\frac{\Theta_D}{1.45 T_c}\right) - 1.04}$$

Assuming a standard value of the Coulomb pseudopotential parameter $\mu^*$ = 0.13 this yields $\lambda_{ep}$ = 0.56 indicative of a weak electron-phonon coupling.

The jump in the specific heat at the superconducting transition is normalized as $\Delta C_p/\gamma T_c$ for a weak-coupling BCS superconductor, and its expected value is 1.43. For MgPd$_2$Sb we obtained $\Delta C_p/\gamma T_c$ =

1.76, indicating a bulk superconducting transition and suggesting moderately coupled superconductivity.

The results of resistivity measurements at high pressures are shown in Fig. 6. A linear decrease in $T_c$ with applied pressure is observed, with the slope of $\Delta T_c/P$ = -0.23(2) K/GPa. A weaker but comparable pressure dependence was observed previously in HfPd$_2$Al [47], where the decrease of $T_c$ was attributed to the effect of lattice stiffening that reduces the electron-phonon coupling coefficient, which is also likely applicable to MgPd$_2$Sb. The reported values of the pressure dependence coefficient of $T_c$ for Heusler compounds are gathered in Table 1.

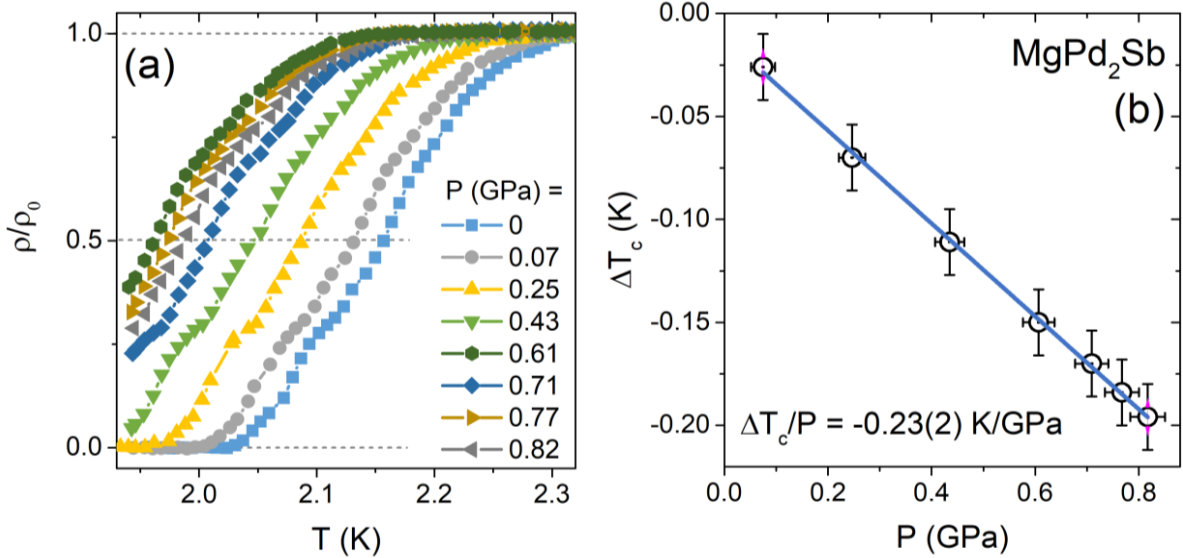

Fig. 6 (a) Plot of temperature-dependent resistivity showing a decrease of $T_c$ with applied pressure. Dashed lines are drawn at $\rho_0$, ½ $\rho_0$ and 0 (b) The change in $T_c$ (as estimated from the ½ resistivity drop criterion) vs. applied pressure. The linear fit (blue line) yields $\Delta T_c/P$ = -0.23(2) K/GPa.

**Theoretical results**

Figure 7 shows the calculated electronic band structure, densities of states (DOS) and Fermi surface of MgPd$_2$Sb. As the constituent atoms of the compound are not particularly heavy, spin-orbit coupling does not visibly influence the electronic states near the Fermi level. Thus further analysis, if not stated otherwise, is based on the scalar-relativistic calculations' results. Two bands cross the Fermi level, which is located on a decreasing slope of the DOS curve, making up two Fermi surface sheets, shown in Fig. 7 (c-d). Total and projected DOS are shown in Fig. 8. The largest contribution to DOS($E_F$) comes from Pd-4d and Sb-5p states. Mg contribution to DOS($E_F$) is negligible. The DOS shape is dictated by Pd states as DOS of Sb and Mg is approximately constant near $E_F$. Total DOS($E_F$) and band-structure value of the Sommerfeld coefficient $\gamma_{band}$ are collected in Table 3. Comparison with the experimental value of $\gamma$ leads to an unexpected situation where $\gamma_{band} \approx \gamma_{expt.}$ i.e. there is no room for the enhancement of $\gamma_{expt}$ by the electron-phonon interaction, as we expect $\gamma_{expt} = \gamma_{band}(1+\lambda_{e-p})$. In this case, $\lambda_{e-p}$ would have to be as small as 0.05, which is completely insufficient to explain superconductivity with $T_c$ = 2.2 K. This suggests that the studied sample must have contained defects and/or disorder, which influenced $\gamma_{expt}$. This scenario is supported by a low RRR value and relatively broad transition to superconducting state.

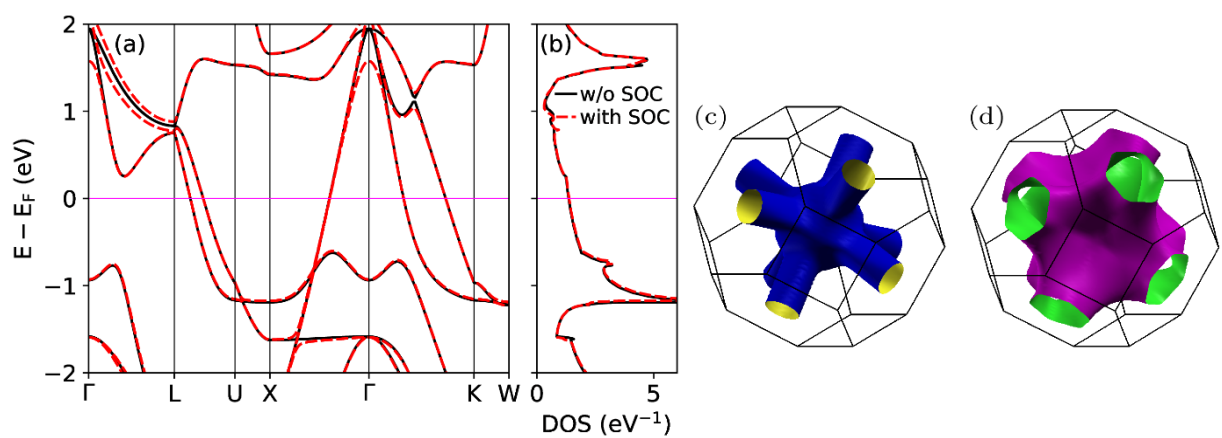

Fig. 7 (a) Electronic band structure and (b) density of states around the Fermi level in MgPd$_2$Sb, calculated with and without spin-orbit coupling; (c,d) two Fermi surface sheets, plotted with XCrysDen [54].

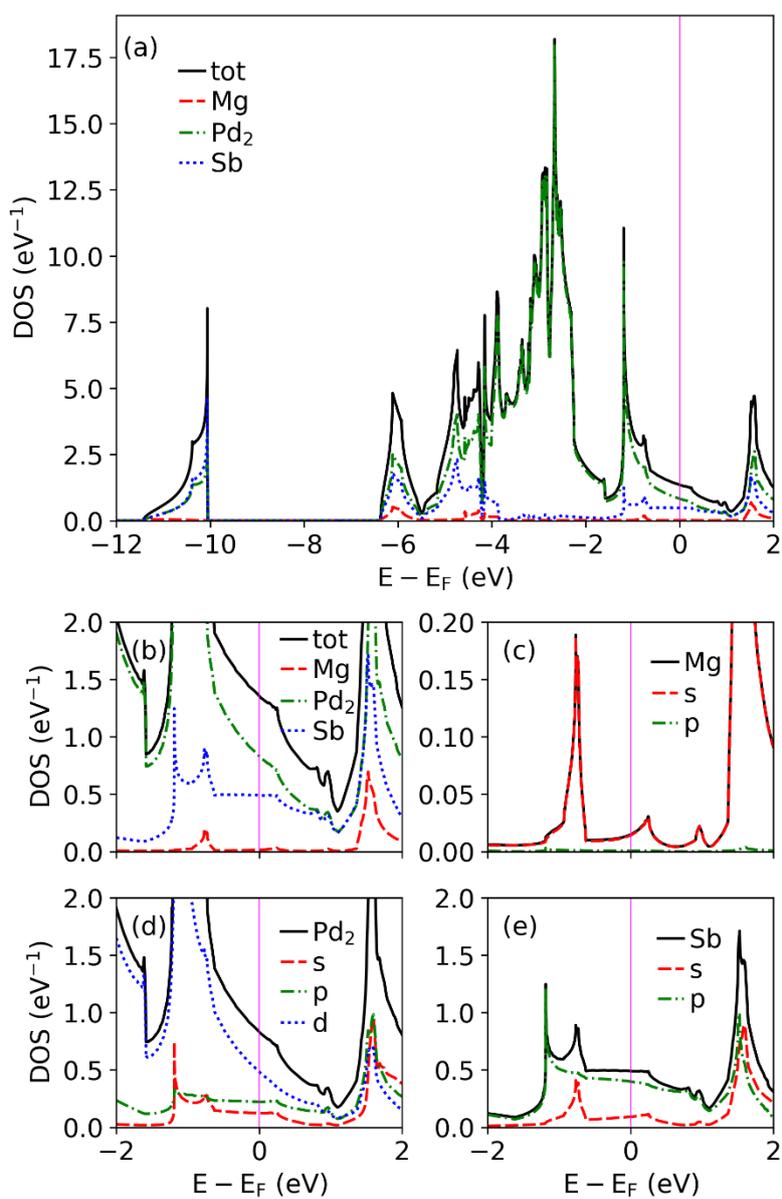

Fig. 8 (a) Total and projected DOS of MgPd$_2$Sb; (b) blow-up close to the Fermi energy; (c)-(e) DOS projected on atomic orbitals.

Table 3 Calculated properties of MgPd$_2$Sb: density of states at Fermi energy DOS($E_F$), bandstructure value of the Sommerfeld coefficient $\gamma_{band}$, electron-phonon coupling parameter $\lambda_{e\text{-}p}$ computed from the Eliashberg function and critical temperature $T_c$ from the Allen-Dynes formula assuming $\mu^* = 0.13$.

|  | DOS($E_F$) (eV$^{-1}$) | $\gamma_{band}$ (mJ mol$^{-1}$ K$^{-2}$) | $\lambda_{e\text{-}p}$ | $T_c$ (K) |
|---|---|---|---|---|
| p = 0 GPa | 1.36 | 3.20 | 0.611 | 2.16 |
| p = 1 GPa | 1.34 | 3.16 | 0.582 | 1.93 |

Phonon dispersion curves and phonon density of states $F(\omega)$ of MgPd$_2$Sb are shown in Fig. 9. The computed phonon structure is stable, as there are no imaginary frequencies and no soft modes, which occur in many Heusler alloys, like in the recently studied LiPd$_2$Ge [43]. As the spin-orbit coupling had negligible effect on the electronic structure of MgPd$_2$Sb, phonon and electron-phonon calculations were done in a scalar-relativistic approximation. We have, however, computed phonons with SOC for two selected **q**-points and calculations confirmed that phonon frequencies were not influenced by SOC, as shown in Figure 10.

Phonon branches of the lightest atom, Mg, are well separated from the heavier Pd and Sb vibrations. It is worth noting that Sb branches are inverted with Pd ones: Sb mass $M_{Sb} = 121.76$u is greater than Pd mass $M_{Pd} = 106.42$ u, but the larger contribution to the lowest acoustic modes is from the lighter Pd atom. Inversion of modes is sometimes proposed to be correlated with the observed dynamical instabilities in some Heusler compounds [40], however such a situation is not seen here. Figure 10 shows phonon dispersion relation with shading proportional to the phonon linewidths $\gamma_{qv}$ which measure the strength of the electron-phonon coupling [55-57]:

$$\gamma_{qv} = 2\pi\omega_{qv} \sum_{ij} \int \frac{d^3k}{\Omega_{BZ}} |g_{qv}(\boldsymbol{k},i,j)|^2 \delta(E_{\boldsymbol{q},i} - E_F) \delta(E_{\boldsymbol{k+q},j} - E_F),$$

$$g_{qv}(\boldsymbol{k},i,j) = \sum_s \left(\frac{\hbar}{2M_s\omega_{qv}}\right)^{1/2} \left\langle \psi_{i,\boldsymbol{k}} \left| \frac{dV_{scf}}{d\hat{u}_{v,s}} \cdot \hat{e}_v \right| \psi_{j,\boldsymbol{k+q}} \right\rangle.$$

$\omega_{qv}$ is the phonon frequency at **q** vector for the mode v, $M_s$ is mass of atom $s$, $\psi_{i,\boldsymbol{k}}$ is an electron wavefunction, $\hat{e}_v$ is a polarization vector and $\frac{dV_{scf}}{d\hat{u}_{v,s}}$ is a change of the electronic potential due to a displacement of the atom $s$ in the direction $u$.

The relatively largest linewidths are observed for the Mg optic modes. Next, the electron-phonon interaction function $\alpha^2F(\omega)$ (Eliashberg function) is calculated by summing contributions from each of the phonon branch:

$$\alpha^2 F(\omega) = \frac{1}{2\pi N(E_F)} \sum_{qv} \delta(\omega - \omega_{qv}) \frac{\gamma_{qv}}{\hbar\omega_{qv}}.$$

The global electron-phonon coupling constant $\lambda$ is calculated from the Eliashberg function:

$$\lambda_{e-p} = 2 \int_0^{\omega_{max}} \frac{\alpha^2 F(\omega)}{\omega} d\omega.$$

and its frequency distribution λ(ω):

$$\lambda(\omega) = 2 \int_0^\omega \frac{\alpha^2 F(\omega')}{\omega'} d\omega'.$$

is plotted on the top of the α²F(ω) function in Fig. 10. As $\lambda_{e-p}$ is inversely proportional to ω² the largest contribution comes from the low-frequency part of the spectrum. The first six, mainly Pd modes, contribute in about 80% to the total $\lambda_{e-p}$ value, leaving 15% and 5% to Sb and Mg -dominated branches. Therefore, electron coupling with Pd atoms vibrations is the most important in superconductivity of MgPd$_2$Sb. The computed value of $\lambda_{e-p}$ = 0.61 is very close to the value estimated from the McMillan formula and experimental T$_c$, in spite of the disagreement of the electronic specific heat coefficients, which suggests that the potential atomic disorder or other effects which influenced the Sommerfeld coefficient value are not important for superconductivity. The theoretical value of the superconducting critical temperature may be now obtained using the Allen-Dynes formula [41]:

$$T_c = \frac{\langle \omega_{log}^{\alpha^2 F} \rangle}{1.20} \exp\left[\frac{-1.04(1 + \lambda_{e-p})}{\lambda_{e-p} - \mu^*(1 + 0.62\lambda_{e-p})}\right],$$

$$\langle \omega_{log}^{\alpha^2 F} \rangle = \exp\left(\int_0^{\omega_{max}} \alpha^2 F(\omega) \ln(\omega) \frac{d\omega}{\omega} \Big/ \int_0^{\omega_{max}} \alpha^2 F(\omega) \frac{d\omega}{\omega}\right).$$

Assuming the coulomb pseudopotential parameter to be of the typical value of μ* = 0.13 we arrive at T$_c$ = 2.16 K, in agreement with the experimental findings. This clearly shows the electron-phonon origin of superconductivity in MgPd$_2$Sb.

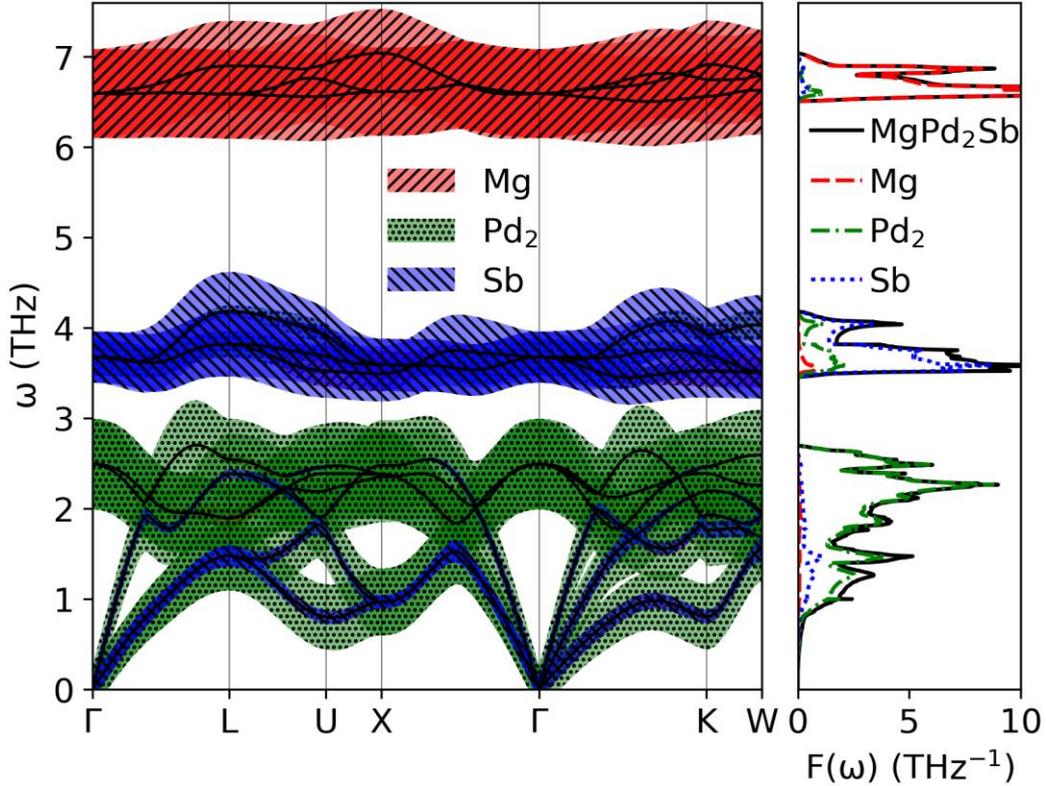

Fig. 9 Phonon dispersion relation and phonon density of states of MgPd$_2$Sb. Color filling with hatches represents atomic contributions to phonon branches.

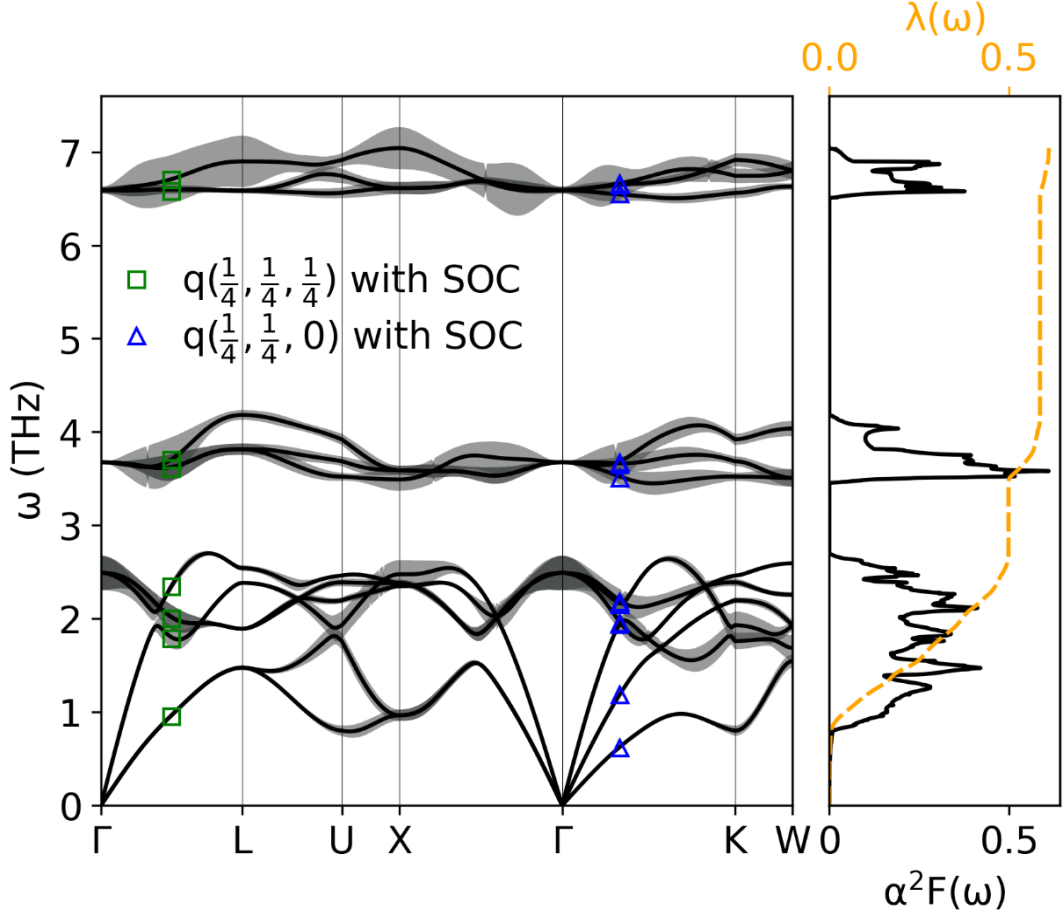

Fig. 10 Left: phonon dispersion relations of MgPd$_2$Sb with shading proportional to phonon linewidths $\gamma_{qv}$. Marked points were calculated with SOC. Right: Eliashberg function (solid line) and cumulative electron-phonon coupling constant (dashed line).

Furthermore, to complete the comparison of the theoretical prediction with experimental works calculations with applied external pressure of 1 GPa were done. The computed lattice parameter is $a$ = 6.528 Å. As this pressure is relatively small, electronic dispersion relations did not change visibly, giving slightly decreased DOS($E_F$) value (see Table 3). However, noticeable differences are observed in the phonon spectrum, presented in Fig. 11. Under the external pressure the Heusler structure remains stable and no phonon softening is observed for the acoustic part of the spectrum. Optical phonon branches move upwards, which is a typical sign of lattice stiffening. The comparison of the Eliashberg functions is additionally done in Fig. 11, and the electron-phonon coupling constant $\lambda_{e\text{-}p}$ visibly decreases from 0.61 (0 GPa) to 0.58 (1 GPa). This is followed by the decrease in the superconducting critical temperature, and the computed rate of $\Delta T_c/p \approx$ -0.23 K/GPa, is again in excellent agreement with the experimental result.

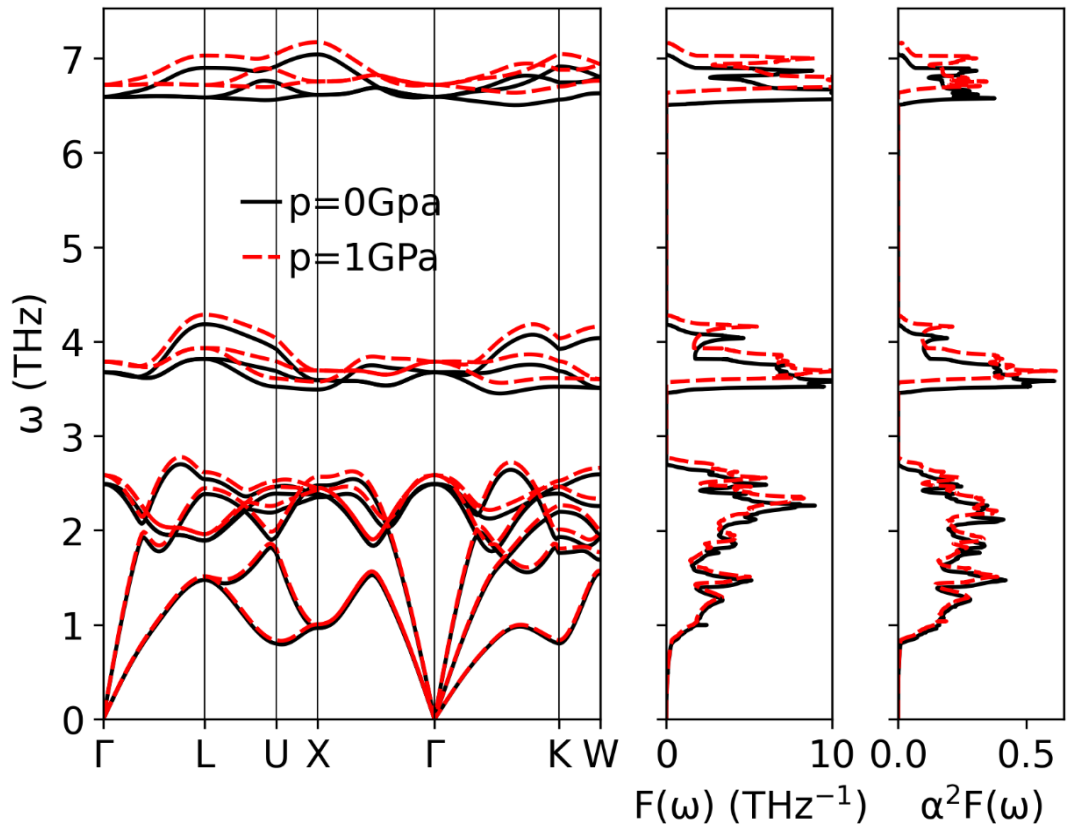

Fig. 11 Effect of pressure on phonon dispersion curves (left panel), phonon density of states $F(\omega)$ (middle panel) and Eliashberg function $\alpha^2 F(\omega)$ (right panel) in MgPd$_2$Sb.

Table 1 Normal and superconducting state properties of Heusler superconductors. Unless noted otherwise the pressure dependence coefficient is taken from ref. [42] and all other parameters from ref. [5] and references therein. Superscripts denote: [a] according to our knowledge LiPd$_2$Ge is the only type-I superconductor in the Heusler family ($\mu_0 H_c$ = 34.2 mT) [43]; [b] $\lambda_{ep}$ was not reported in the literature and was calculated using the McMillan equation based on the reported values of Debye temperature and $T_c$; [c] extrapolated $\Theta_D$ values, likely underestimated; [d] $\xi_{GL}$ calculated based on the reported value of the upper critical field. Numbers in parentheses are uncertainties of the least significant digits (where available); * - superconductivity coexistent with a charge density wave distortion, † - superconductivity coexistent with magnetic order.

| Property | $T_c$ (K) | $\Theta_D$ (K) | $\gamma$ (mJ mol$^{-1}$K$^{-2}$) | $\Delta C_p/\gamma T_c$ | $\lambda_{ep}$ | $\mu_0 H_{c2}(0)$ (T) | $\xi_{GL}$ (nm) | $\Delta T_c/P$ (K/GPa) |
|---|---|---|---|---|---|---|---|---|
| VEC = 16 | | | | | | | | |
| LiGa$_2$Rh | 2.4[44] | 320[44] | 4.7[44] | 1.48[44] | 0.52[44] | 0.31[44] | 33[44] | - |
| VEC = 25 | | | | | | | | |
| LiPd$_2$Ge | 1.96[43] | 194(3)[43] | 5.8(1)[43] | 1.38[43] | 0.56[43] | [a] | - | - |
| VEC = 26 | | | | | | | | |
| YPd$_2$In | 0.85 | - | - | - | - | - | - | - |
| LuPt$_2$In * | 0.45 [45] | 190 [45] | - | - | 0.43 [b] | - | - | - |
| VEC = 27 | | | | | | | | |
| **MgPd$_2$Sb** | **2.1** | **201(2)** | **3.35(8)** | **1.76** | **0.56** | **0.194(3)** | **41.2(3)** | **-0.23(2)** |
| ZrPd$_2$Al | 3.4 | 189(1) | 9.0(1) | 1.02 | 0.65 | 2.82 | 11 | - |
| HfPd$_2$Al | 3.66 | 182(3) | 7.9(3) | 1.50 | 0.68 | 1.81 | 13 | -0.13(1) [46] |
| ZrPd$_2$In | 2.19 | 236(5) | 10.9(2) | - | 0.55 | 0.63 | 23 | - |
| HfPd$_2$In | 2.86 | 243(5) | 8.5(2) | 1.72 | 0.58 | 1.00 | 18 | - |
| ScPd$_2$Sn | 2.0 | 277(1) | 6.6(2) | - | 0.52 | 0.26 | 36 | -0.15(10) |
| YPd$_2$Sn | 4.7 | 210(4) | 9.2(2) | 1.73 | 0.70 | 0.90 | 19 | -0.226(21) |
| LuPd$_2$Sn | 2.8 | 246(2) | 7.4(1) | 1.45 | 0.58 | 0.45 | 27 | -0.168(7) |
| ScPd$_2$Pb | 2.4 | - | - | - | - | - | - | - |
| YPd$_2$Pb | 2.3 | 198 [47] | - | - | 0.58 [b] | | | -0.193(9) |
| LuPd$_2$Pb | 2.4 | - | - | - | - | - | - | - |
| ErPd$_2$Sn † | 1.17 | - | - | - | - | - | - | - |
| TmPd$_2$Sn | 2.82 | 120 [c] [47] | - | | 0.72 [b] | | | - |
| YbPd$_2$Sn † | 2.46 | 118 [c] [47] | 6.0 [48] | | 0.69 [b] | | | -0.193(9) |
| TmPd$_2$Pb | 2.1 | - | - | - | - | - | - | -0.262(9) |
| YbPd$_2$Pb | 2.8 | - | - | - | - | - | - | - |
| ZrNi$_2$Al | 1.38 | 278(2) [49] | 13.7(2)[49] | - | 0.48 [b] | - | - | - |
| HfNi$_2$Al | 0.74 | 288(1) [49] | 10.9(2)[49] | - | 0.43 [b] | - | - | - |
| ZrNi$_2$Ga | 2.85[50] | 300[50] | 17.3[50] | 1.41[50] | 0.55 [50] | 1.48[50] | 15 [d] | - |
| HfNi$_2$Ga | 1.12 | - | - | - | - | - | - | - |
| ScPdAuAl | 3.0[51,52] | - | - | - | - | - | - | - |
| ScPtAuIn | 0.96[51] | - | - | - | - | - | - | - |
| YPdAuIn | 2.6[51] | - | - | - | - | - | - | - |
| VEC = 28 | | | | | | | | |
| ScAu$_2$Al | 4.4 | - | - | - | - | - | - | - |
| NbNi$_2$Al | 2.15 | 280 [53] | 8.0 [53] | 1.8 [53] | 0.52 [53] | 1.35 [53] | 16 [d] | - |
| NbNi$_2$Ga | 1.54 | 240 [53] | 6.5[53] | - | 0.50[53] | 0.60[53] | 23 [d] | - |
| ScAu$_2$In | 3 | - | - | - | - | - | - | - |
| YAu$_2$In | 1.74 | - | - | - | - | - | - | - |
| YPd$_2$Sb | 0.85 | - | - | - | - | - | - | - |
| VEC = 29 | | | | | | | | |
| NbNi$_2$Sn | 2.90 | 206[53] | 4.0[53] | 1.6[53] | 0.61[53] | 0.63[53] | 23 [d] | - |

## Conclusions

As noted earlier [5,51], most of the known superconductors show the VEC = 27. This was recently rationalized by Kautzsch *et al*. [51] based on rigid band considerations of the electronic structure of Heusler compounds: at a certain electron count VEC ≈ 27 the Fermi level lies in the proximity of a van Hove singularity, resulting in an increased electronic density of states (DOS) and a tendency to either undergo a structural distortion (as in the case of LuPt$_2$In [45]) or superconducting transition. The $T_c$ differences between isoelectronic compounds are then a result of subtle differences in electronic structure, phonon structures and/or the amount of structural disorder.

Based on electron counting rules and supported by results of high-throughput calculations we have synthesized a previously unreported, Mg-bearing Heusler compound, which, in accordance with the theoretical calculations, was found to be superconducting, with $T_c \approx 2.2$ K. MgPd$_2$Sb is a weakly coupled type-II BCS superconductor.

Table 1 gathers normal and superconducting state parameters for MgPd$_2$Sb and other reported full Heusler superconductors. Compared to other members of this family for which the data are available, MgPd$_2$Sb has the lowest value of the electronic heat capacity coefficient γ and the upper critical field $H_{c2}$. The electron-phonon coupling coefficient $\lambda_{ep}$ is close to observed in other compounds with similar $T_c$ (ZrPd$_2$In, HfPd$_2$In, YPd$_2$Pb, NbNi$_2$Al, LiGa$_2$Rh), which is consistent with the McMillan formula, as most of the Heusler superconductors have similar Debye temperature on the order of 200 K. The pressure dependence of $T_c$ was found to be second strongest among the reported values after TmPd$_2$Pb.

## Acknowledgements

Work at GUT was supported by the National Science Centre (Poland), Grant No. 2017/27/B/ST5/03044. Work at AGH-UST was supported by the National Science Centre (Poland), Project No. 2017/26/E/ST3/00119. The synthetic work at Princeton was supported by the US Department of Energy, Division of Basic Energy Sciences, Grant No. DE-FG02- 98ER45706.